\documentclass[final,5p,times,twocolumn]{elsarticle}
\usepackage{graphicx,float,amssymb,xcolor,url,ulem,enumitem}
\biboptions{sort&compress}
\usepackage{lipsum,,mathrsfs}
\usepackage{hyperref}
\usepackage[utf8]{inputenc}
\usepackage[T1]{fontenc}

\journal{Physics Letters B}
\begin{document}
\begin{frontmatter}
\title{First investigation on the isomeric ratio in multinucleon transfer reactions: Entrance channel effects on the spin distribution}
\author[first]{D. Kumar\corref{cor1}}
\ead{De.Kumar@gsi.de, dm978dph@gmail.com}
\cortext[cor1]{Corresponding author}
\author[first,second]{T. Dickel}
\author[second,third]{A. Zadvornaya}
\author[third]{O. Beliuskina}
\author[third]{A. Kankainen}
\author[forth]{P.~Constantin}
\author[first]{S.~Purushothaman}
\author[forth]{A.~Spataru}
\author[third]{M.~Stryjczyk}
\author[third,fifth]{L.~Al~Ayoubi}
\author[sixth]{M.~Brunet}
\author[third]{L.~Canete}
\author[third,fifth]{C.~Delafosse}
\author[third,seventh]{R.P.~de~Groote}
\author[third]{A.~de~Roubin}
\author[third]{T.~Eronen}
\author[first,third]{Z.~Ge}
\author[third]{W.~Gins}
\author[first]{C.~Hornung}
\author[third,eighth]{M.~Hukkanen}
\author[third]{A.~Illana~Sison}
\author[third]{A.~Jokinen}
\author[forth]{D.~Kahl}
\author[first]{B.~Kindler}
\author[first]{B.~Lommel}
\author[ninth,tenth]{I.~Mardor}
\author[third]{I.D.~Moore}
\author[third]{D.A.~Nesterenko}
\author[forth]{D.~Nichita}
\author[third]{S.~Nikas}
\author[third]{A.~Ortiz-Cortes}
\author[third]{H.~Penttilä}
\author[sixth]{Zs.~Podoly\'{a}k}
\author[third]{I.~Pohjalainen}
\author[third]{A.~Raggio}
\author[third]{M.~Reponen}
\author[third]{S.~Rinta-Antila}
\author[third,eleventh]{J.~Romero}
\author[third]{M.~Vilen}
\author[third]{V.~Virtanen}
\author[twelveth]{A.~Weaver}
\author[first]{J.~Winfield\fnref{fn1}}
\fntext[fn1]{Deceased.}
\affiliation[first]{city={GSI Helmholtzzentrum f\"ur Schwerionenforschung GmbH, Darmstadt}, postcode={64291}, country={Germany}}    
\affiliation[second]{city={II. Physikalisches Institut, Justus-Liebig-Universit\"at Gie\ss{}en, Gie\ss{}en}, postcode={35392}, country={Germany}}
\affiliation[third]{city={University of Jyv\"askyl\"a, Department of Physics, Accelerator laboratory}, postcode={P.O. Box 35(YFL)}, country={FI-40014 University of Jyv\"askyl\"a, Finland}}
\affiliation[forth]{city={Extreme Light Infrastructure-Nuclear Physics, Horia Hulubei National Institute for R\&D in Physics and Nuclear Engineering, M\u{a}gurele}, postcode={077125}, country={Romania}}
\affiliation[fifth]{city={Universit\'e Paris Saclay, CNRS/IN2P3, IJCLab}, postcode={91405}, country={Orsay, France}}
\affiliation[sixth]{city={Department of Physics, University of Surrey, Guildford}, postcode={GU2 7XH}, country={United Kingdom}}
\affiliation[seventh]{city={KU Leuven, Instituut voor Kern- en Stralingsfysica, Leuven}, postcode={B-3001}, country={Belgium}}
\affiliation[eighth]{city={Universit\'e de Bordeaux, CNRS/IN2P3, LP2I Bordeaux}, postcode={UMR 5797, F-33170}, country={Gradignan, France}}
\affiliation[ninth]{city={Tel Aviv University, Tel Aviv},postcode={6997801}, country={Israel}}
\affiliation[tenth]{city={Soreq Nuclear Research Center, Yavne}, postcode={8180000}, country={Israel}}
\affiliation[eleventh]{city={Oliver Lodge Laboratory, University of Liverpool, Liverpool}, postcode={L69 7ZE}, country={UK}}
\affiliation[twelveth]{city={School of Computing, Engineering and Mathematics, University of Brighton, Brighton}, postcode={BN2~4GJ}, country={United Kingdom}}

\begin{abstract}
The multinucleon transfer (MNT) reaction approach was successfully employed for the first time to measure the isomeric ratios (IRs) of $^{211}$Po (25/2$^+$) isomer and its (9/2$^+$) ground state at the IGISOL facility using a 945 MeV $^{136}$Xe beam impinged on $^{209}$Bi and $^{\rm nat}$Pb targets.\ The dominant production of isomers compared to the corresponding ground states was consistently revealed in the $\alpha$-decay spectra.\ Deduced IR of $^{211}$Po populated through the $^{136}$Xe+$^{\rm nat}$Pb reaction was found to enhance $\approx$1.8-times than observed for $^{136}$Xe+$^{209}$Bi.\ State-of-the-art Langevin-type model calculations have been utilized to estimate the spin distribution of an MNT residue.\ The computations qualitatively corroborate with the considerable increase in IRs of $^{211}$Po produced from $^{136}$Xe+$^{\rm nat}$Pb compared to $^{136}$Xe+$^{209}$Bi.\ Theoretical investigations indicate a weak influence of target spin on IRs.\ The enhancement of the $^{211}$Po isomer in the $^{136}$Xe+$^{\rm nat}$Pb over $^{136}$Xe+$^{209}$Bi can be attributed to the different proton ($p$)-transfer production routes.\ Estimations demonstrate an increment in the angular momentum transfer, favorable for isomer production, with increasing projectile energy.\ Comparative analysis indicates the two entrance channel parameters, projectile mass and $p$-transfer channels, strongly influencing the population of the high-spin isomer of $^{211}$Po (25/2$^+$).\ This work is the first experimental and theoretical study on the IRs of nuclei produced via different channels of MNT reactions, with the latter quantitatively underestimating the former by a factor of two.\
\end{abstract}
\end{frontmatter}

\section{\label{s1}Introduction}
The crux of exploring nuclear reaction and structural properties of heavy neutron-rich nuclei is to grasp an understanding of the evolution of shell structure far from the valley of $\beta$-stability, which in turn is crucial for the astrophysical rapid neutron capture process ($r$-process) \cite{Otsuka19,Horowitz19,Heinz22}.\ Of all the nuclear properties, experimental data on isomeric ratios (IRs) of reaction products is essential to comprehend the spin distributions, which eventually affect the population of isomers.\ Knowledge of the proper production routes of exotic neutron-rich isomers is vital for studying nuclear structural aspects like half-lives, spins, decay paths, etc., which may play a prominent role in driving the $r$-process pathways \cite{Misch21,Ani05}.\ Moreover, nuclear isomers can be exploited for applications like nuclear medicine, energy storage, etc.\ \cite{DMKumar21,Maiti17,DMKumar17,DMKumar19,Walker99}.\

The MNT reaction approach has emerged as a pragmatic pathway of accessing heavy neutron-rich uncharted terrain of the nuclear chart spanning from the rare-earth region to the "Island of Stability" \cite{Zhu22,DeKumar21,Vidal23}.\
The MNT fragments, heavier than fission-like fragments, are extended across quasi-elastic (QE) and deep-inelastic collision (DIC) regimes \cite{Vidal23,Kumar21,Novikov20,Kozulin23,Kumar20}.\ 
The promising outcomes of advanced MNT models at neutron shell closures, particularly $N$=126, have steered many experiments towards using $^{136}$Xe+$^{208}$Pb, $^{136}$Xe+$^{198}$Pt, $^{204}$Hg+$^{208}$Pb, and $^{64}$Ni+$^{207,208}$Pb reactions at above barrier energies during the last two decades\cite{Zagrebaev08,Karpov17,Saiko19,Saiko22,Li16,Li18,Li20,Zhu21,Zhao21,Kozulin12,Watanabe15,Utepov23,Son23,Clement23,Welsh17,Desai20EPJA,VDesai20,Barrett15,Comas13,Krolas03}.\ However, none of them were focused on studies of spin distribution of MNT fragments.\ To date, comprehensive studies of MNT reaction properties for limited target-projectile systems have been carried out using in-flight electromagnetic separators such as VAMOS++ along with EXOGAM, AGATA, and CATLIFE at GANIL \cite{Watanabe15,Utepov23,Son23,Clement23}, PRISMA coupled to AGATA and CLARA at INFN LNL in Legnaro \cite{Vidal23,Gadea04,Vogt15,Dobon23}, MAGNEX at INFN-LNS in Catania \cite{Cappuzzello16,Koulouris23}, SHIP velocity filter at GSI \cite{Heinz22,Comas13}, including decay/in-beam $\gamma$-ray spectroscopic method using Gammasphere facility at ANL \cite{Welsh17,Desai20EPJA,VDesai20,Barrett15}.\ However, the discovery of any unknown nuclei would be precluded in many of the aforementioned methods, in which the identification of nuclei can only be determined by known $\gamma$- or $\alpha$-decay patterns.\
In addition, similar studies were also carried out using the CORSET setup at Dubna \cite{Kozulin12}, $\Delta$$E$–$E$ Si-detector telescope together with four MWPCs at JAEA \cite{Hirose17}.\

The strenuous attempt to overcome experimental challenges in the production, separation, and identification of heavy MNT fragments is still being continued across various nuclear laboratories around the globe \cite{Loveland07,Chen20,Timo20}.\ Several ion-catcher setups have been commissioning in recent years to span a broader coverage of angular distribution of MNT products (e.g., $\approx$20$^{\circ}$-60$^{\circ}$, \cite{Liao23}), such as MNT-ion gas cell at IGISOL \cite{Ana20,SashaNIMB}, FRS ion-catcher (IC) with INCREASE at GSI \cite{Rotaru22,Timo20}, $N$=126 factory at ANL \cite{Savard20}, and NEXT at Groningen \cite{Even22}.\
The KISS experiment at Riken has recently led to breakthroughs by measuring an unknown uranium isotope, $^{241}$U, together with many neutron-rich projectile-like fragments (PLFs) of Pa-Am isotopes in an MNT reaction of $^{238}$U+$^{198}$Pt \cite{Niwase23}.\ Moreover, spectroscopic investigations on several target-like fragments (TLFs) of Os-Pt isotopes in many experiments of $^{136}$Xe+$^{198}$Pt have built up promising prospects for the MNT methodology to produce many more undiscovered nuclei \cite{Hirayama17,Hirayama18,Mukai20,Choi20}.\
In this endeavor, a comprehensive and systematic program for MNT reactions using newly developed different types of MNT-ion catchers at IGISOL (JYFL) and FRS-IC (GSI) facilities have been initiated in the interest of addressing both aspects of nuclear features:\ (i) to disentangle the MNT reaction processes in addition to benchmarking state-of-the-art MNT models; and (ii) to produce neutron-rich exotic isotopes and isomers for nuclear structural studies and mass measurements \cite{Timo20,Plass19,TimoD16,Ana20,Raggio23,Dima23,Jaries23}.\ 

Recent MNT models differ by orders of magnitude for the nuclei produced by transferring a few nucleons from the target/projectile (e.g., $\Delta$$Z$ $\ge$ 2) or for symmetric target-projectile systems \cite{Desai20EPJA,VDesai20,Barrett15,Welsh17}.\ This generates urgent interest in validating these models by comparing them with the reaction data of different forms.\ This letter reports the first measurements of IRs populated via the MNT reactions.\ The IR manifests the characteristics of the spin distribution of MNT fragments.\ Moreover, computations of the spin distribution of the MNT fragment are performed for the first time using improved state-of-the-art Langevin-type model calculations and benchmarked by comparing it with the measured IRs.\

\section{\label{s2} Experimental study}
A series of experiments were performed during the commissioning of dedicated MNT gas cells with the aim of accessing neutron-rich exotic nuclei utilizing the MNT approach at the Ion-Guide Isotope Separator On-Line (IGISOL) facility of the JYFL Accelerator Laboratory, University of Jyväskylä, Finland \cite{Arje86,Moore13}.\ The TLFs were produced using the 945 MeV $^{136}$Xe$^{31+}$ beam delivered by the K-130 heavy-ion cyclotron, impinged on $\approx$5.1 mg/cm$^2$ $^{209}$Bi and $\approx$6 mg/cm$^2$ $^{\rm nat}$Pb targets.\ The projectile energy loss within the target was estimated as 945--792 MeV for $^{209}$Bi and 945--765 MeV for $^{\rm nat}$Pb using the SRIM code \cite{Ziegler10}, which results in the production yield of $^{211}$Po at mid-target energy, 868$\pm$77 MeV (or equivalently, energy above the Coulomb barrier, E/V$_{\rm B}$=1.23$\pm$0.11) and 855$\pm$90 MeV (E/V$_{\rm B}$=1.21$\pm$0.12), respectively.\

\begin{figure}[t]
	\centering
	\includegraphics[width=88mm,height=48mm]{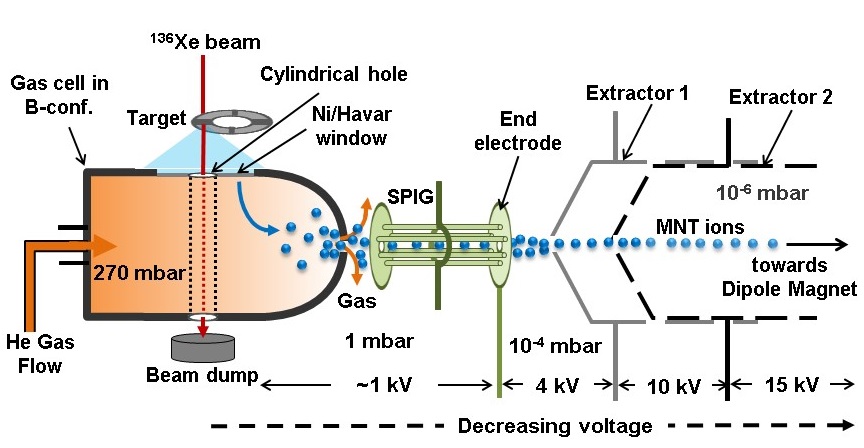}
	\caption{\label{fig1}(Color online)\ A schematic diagram of the experimental arrangements consisting of a target, the MNT gas cell in B-configuration \cite{SashaNIMB}, the beam dump, He gas flow, the sextupole ion guide (SPIG), and the extractor electrodes \cite{Karv08}.}
\end{figure}	
\begin{figure}[t]
	\centering
	\includegraphics[width=84mm,height=153mm]{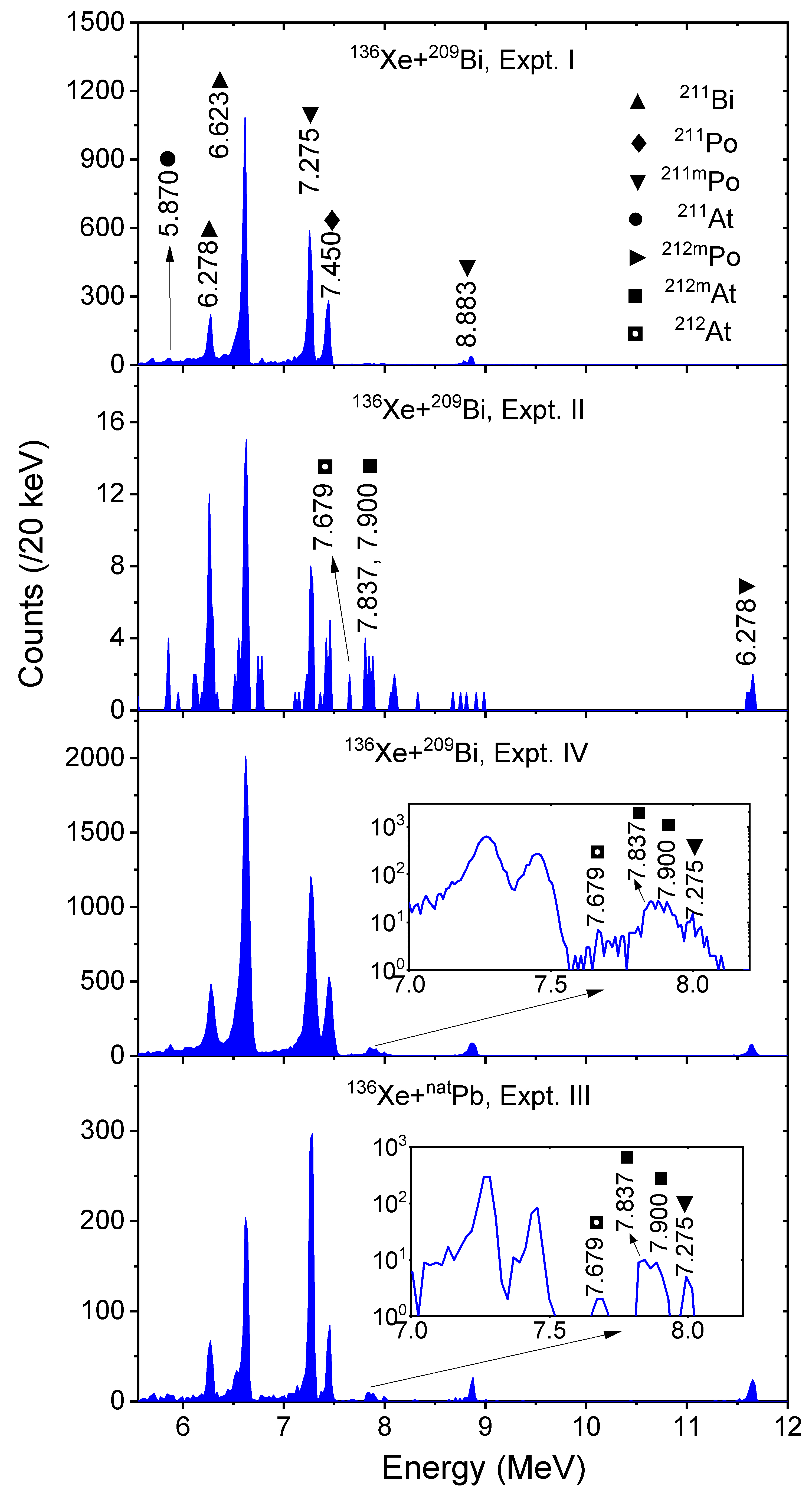}
	\caption{\label{fig2}(Color online)\ Characteristics $\alpha$-decay spectra of TLFs measured in three independent experiments using $^{136}$Xe+$^{209}$Bi at E/V${\rm _B}$ = 1.23$\pm$0.11 and one with $^{136}$Xe+$^{\rm nat}$Pb at E/V${\rm _B}$ = 1.21$\pm$0.12 for different gas cell configurations.}
\end{figure}

A schematic diagram of the experimental setup consists of a target, the gas cell, the beam dump, the sextupole ion guide (SPIG), and the extractor electrodes shown in Fig.\ \ref{fig1}.\
The energetic MNT fragments produced from the target were traversed through a nickel(Ni) or havar window of the gas cell.\ Different configurations of the gas cells (Modified HIGISOL, MNT gas cell in A- and B-configuration) were used and tested in different experiments \cite{SashaNIMB}.\ The primary beam was stopped within a graphite beam-dump mounted either in front of the gas cell (modified HIGISOL and MNT gas cell in A-configuration) or after the gas cell (MNT gas cell in B-configuration) \cite{SashaNIMB}.\ The MNT products were stopped within the He buffer gas inside the gas cell.\ The gas flow subsequently extracted the thermalized ions from the gas cell, from which they were guided through a radiofrequency SPIG and extractor electrode system towards the mass separator\cite{Karv08}.\
The target chamber was kept at +30 kV.\ The extracted ions were gradually accelerated towards the grounded electrostatic switchyard (SW), with typical voltage differences shown in Fig.\ \ref{fig1} and mass separated using the dipole magnet having a mass resolving power ($M/$$\Delta$$M$), $R$ $\approx$ 300, placed before the SW.\ Finally, in-beam $\alpha$-decay spectra were measured using a Silicon (Si) detector mounted at the SW.\

Four different experiments were performed to measure MNT residues between June 2019 -- November 2021, as shown in Fig.\ \ref{fig2}.\
The characteristic $\alpha$-peaks of $^{211}$Bi, $^{\rm 211m}$Po, $^{211}$Po, and $^{\rm 212m}$Po were clearly identified in addition to a minute amount of $^{211}$At, $^{\rm 212m}$At, and $^{212}$At.\ The dipole magnet setting was at mass, $A$ = 211.\
The presence of $\alpha$-peaks from the neighboring mass, $A$ = 212, is due to a limited resolving power of the dipole magnet.\
As evident from Fig.\ \ref{fig2}, the relative production of different $\alpha$-emitting MNT products is found to be consistent across spectra, with $^{211}$Bi as the dominant peak in all three measurements of $^{136}$Xe+$^{209}$Bi reaction.\
However, the most intense $\alpha$-decay peak of $^{211}$Po (i.e., 7225 keV) was dominantly observed in the $^{136}$Xe+$^{\rm nat}$Pb reaction among all other peaks.\ This clearly reveals the relative enhanced production of $^{211}$Po isomer over its ground-state in the $^{136}$Xe+$^{\rm nat}$Pb than the $^{136}$Xe+$^{209}$Bi.\ 
It is important to bear in mind that the broader angular distributions of MNT fragments would result in a wider distribution of the stopping position of ions inside the gas cell \cite{Karpov17}.\ The wider distribution would cause a significant variation in extraction time, i.e., transport time from the gas cell to the SW (typically 100 ms).\ This leads to a larger uncertainty in the yields of $^{\rm 212m}$At (0.119 s) and $^{\rm 212}$At (0.314 s) in addition to the lower statistics.\ Therefore, the estimation of IRs of $^{211}$At is excluded.\

The IR of $^{211}$Po was scrutinized against various experimental conditions such as:\ (i) different beam intensities, $I$ (pnA); (ii) angular coverage of the gas cell window for the MNT products; (iii) He gas pressure within the gas cell, $P$ (mbar); and have been tabulated in Table \ref{tab2}.\ Additionally, the carbon (C) catcher foils before the Si detector, the slit width opening at the electrostatic SW, and the thickness of the gas cell window, including the used material, would also affect the product yields and are, therefore, enlisted in Table \ref{tab2}.\
The isomeric yield (i.e., isomer-to-ground state) ratio (IYR) of the $^{211}$Po was deduced and normalized with the corresponding $\alpha$-peak intensities.\ Therefore, the deduced IYR would be equivalent to the isomeric cross-section ratio (ICR), independent of experimental parameters such as manifested in Table \ref{tab2} as well.\ Hence, IYR or ICR are referred to as IR in this letter.\

\begin{table*}[t]
	\caption{Measured yields (Y) of $^{\rm 211m}$Po, $^{211}$Po, and the subsequently deduced isomeric ratio (IR) of $^{211}$Po for a 945 MeV $^{136}$Xe$^{31+}$ beam at different experimental conditions: intensity ($I$) of the beam, angular coverage (Ang.\ cov.) of the gas-cell window, pressure ($P$) of the He gas, thickness of carbon (C) foil placed before the Si detector, and slit width opening at the entrance to the electrostatic switchyard. The thickness of the entrance window foil of the gas cell was 4.3 mg/cm$^2$ Havar used in Expt.\ I-II, whereas 4.4 mg/cm$^2$ Ni was used in Expt.\ III-IV.\
	}
	\label{tab2}
		\begin{center}
			\renewcommand{\arraystretch}{0.96}
			\begin{tabular}{ccccccccccc}
				\hline
				Expt.\ & Month/Year & $I$  & Ang.\ cov.  & $P$ & C foil & width  &  Y$_{^{\rm 211m}\mathrm{Po}}$ & Y$_{^{211}\mathrm{Po}}$  & IR of \\
			& (Target)	 & (pnA) & (deg.\ ($^{\circ}$))  & (mbar) & ($\mu$g/cm$^2$) & (mm) &  (1/min) &  (1/min) & $^{211}$Po\\ \hline
				Expt. I & 06/2019 & 10 & 14--35 & 300 &	- &	7 &	 8.5 $\pm$ 0.2 &	4.0 $\pm$ 0.2 & 2.3 $\pm$ 0.2 \\
				& ($^{209}$Bi) & 10 & 14--35 & 300 &	- &	7 &	 8.9 $\pm$ 0.2 &	4.1 $\pm$ 0.2 & 2.4 $\pm$ 0.2 \\ \hline
				Expt. II & 08/2019 & 20 & 14--35 & 300 &	- &	10 &  1.4 $\pm$ 0.2 &	0.8 $\pm$ 0.1  & 1.9 $\pm$ 0.5 \\
				& ($^{209}$Bi) & 20 & 11--31 & 300 &	- &	10 & 0.7 $\pm$ 0.1 & 0.4 $\pm$ 0.1 & 1.8 $\pm$ 0.7 \\ \hline
				Expt. III & 03/2021 & 30  & 27--60 & 235 & - & 7.5 &  8.9 $\pm$ 0.3 & 2.0 $\pm$ 0.2 & 4.8 $\pm$ 0.5 \\
				& ($^{\rm nat}$Pb) & 30 & 27--60& 250 & 195 & 7.5 &  6.4 $\pm$ 0.3 & 1.7 $\pm$ 0.2  & 4.0 $\pm$ 0.6 \\
				& & 20 & 45--65& 220 & 190 & 7.5 & 2.0 $\pm$ 0.1 & 0.5 $\pm$ 0.1  & 4.7 $\pm$	1.0 \\
				& & 20 & 27--60 & 220 & 190 & 7.5 & 4.9 $\pm$ 0.2 &	1.2 $\pm$ 0.1  & 4.4 $\pm$	0.6 \\
				& & 30 & 27--60 & 270 &	195 & 7.5 & 5.4 $\pm$ 0.2 & 1.3 $\pm$ 0.1  & 4.4 $\pm$	0.6 \\ \hline
				Expt. IV & 11/2021 & 30 & 20--55 & 265 &	- &	10 & 75.4 $\pm$	1.5 & 32.4 $\pm$ 1.0 & 2.5 $\pm$	0.1 \\
				& ($^{209}$Bi) & 20 & 17--51 & 260 &	205 & 10 & 28.4 $\pm$ 1.1 &	11.8 $\pm$	0.7 &	2.6 $\pm$ 0.2 \\
				& & 28 & 17--51 & 262 &	- &	10 & 73.8 $\pm$	1.6 & 29.3 $\pm$ 1.0  & 2.7 $\pm$	0.2 \\
				& & 20 & 14--48 & 262 &	195 & 10 & 42.2 $\pm$ 1.3 &	18.8 $\pm$ 0.9 &	2.4 $\pm$ 0.2 \\
				& & 31 & 17--51 & 270 &	- &	10 & 85.3 $\pm$	1.2 & 34.9$\pm$	0.8 & 2.7$\pm$	0.1 \\ \hline
			\end{tabular}
		\end{center}
\end{table*}


\section{\label{s3} Computations of spin distributions and IRs}
\begin{figure}[t]
	\centering
	\includegraphics[width=84mm]{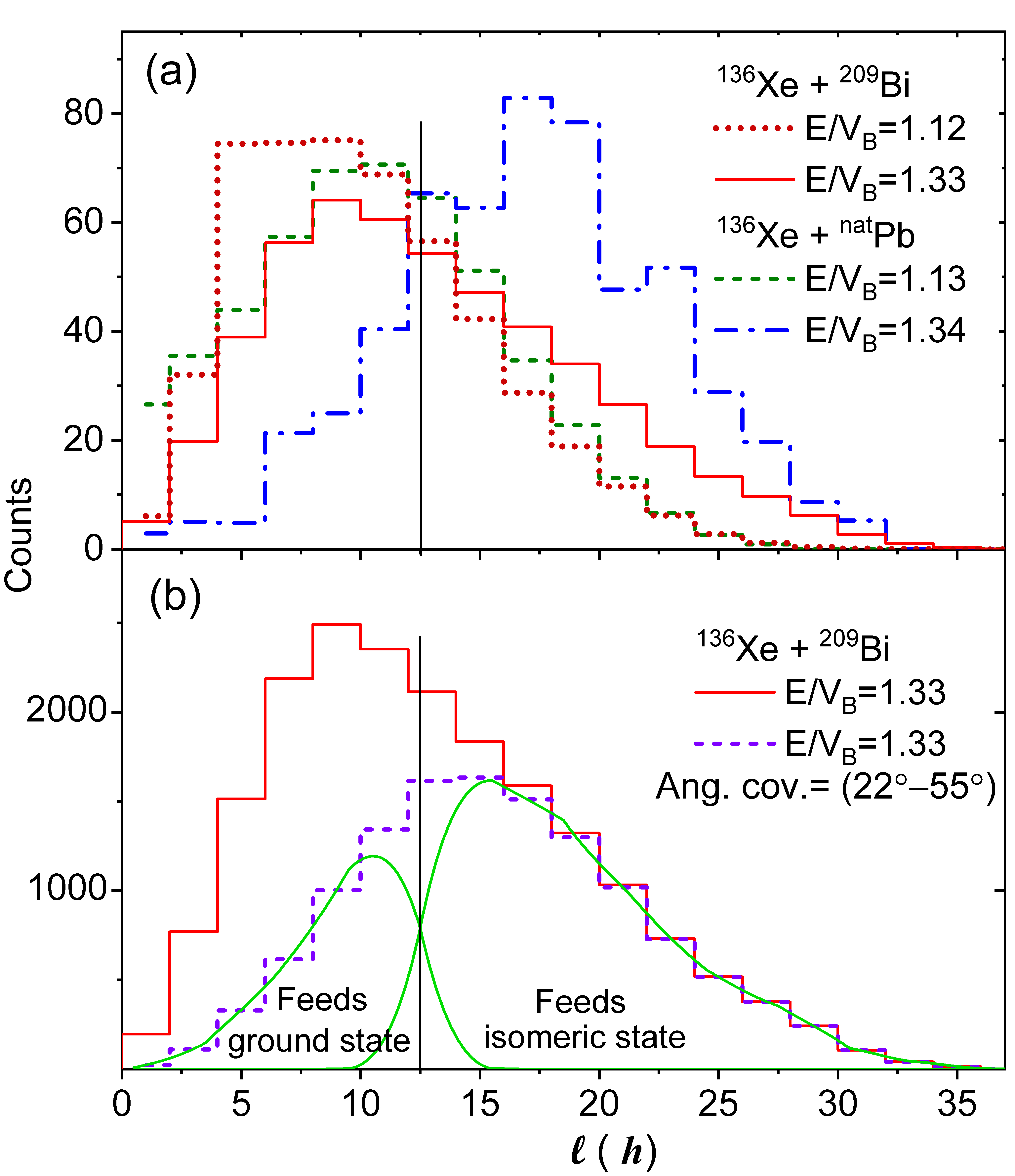}	
	\caption{\label{fig3}(Color online)\ (a) Spin distribution of $^{211}$Po using a dynamical model Langevin approach for $^{136}$Xe+$^{209}$Bi at 12\% (dotted line) and 33\% (solid line) above the Coulomb barrier, and for $^{136}$Xe+$^{\rm nat}$Pb at 13\% (dashed line) and 34\% (dashed-dotted line) above barrier.\ (b) Unfolding of $^{211}$Po into the ground and isomeric states corresponding to 22$^{\circ}$-55$^{\circ}$ of angular coverage of gas cell window. The vertical line indicates the spin value of $^{\rm 211m}$Po (25/2$^+$).}
\end{figure}
A multidimensional dynamical approach based on Langevin equations has been adopted to examine the measured data for the $^{136}$Xe+$^{209}$Bi/$^{\rm nat}$Pb.\ The dynamical approach has adequately reproduced the mass, charge, energy, and angular distributions of the MNT-induced products for most of the limited studied reactions so far \cite{Zagrebaev08,Karpov17,Saiko19,Saiko22}.\ The dynamical model was expanded to provide information about the spin distribution, thereby enabling the theoretical study of the spin distribution of MNT reaction products for the first time.\
The calculation of IR of a nuclide produced via MNT reaction can be conceptualized into two steps:\ (i) estimation of the spin distribution of an MNT product, and (ii) feeding of isomeric and ground states from the spin distribution.\
In the first step, the total angular momentum distribution was estimated by folding the orbital angular momentum brought in by the projectile with the non-zero intrinsic spin of the target.\ Moreover, the exchange of angular momentum due to the transfer of nucleons from projectile to target is necessary for the population of trans-target products, e.g., $^{211}$Po, $^{211}$Bi, etc., and was considered in the calculation.\ This was implemented as the sequential transfer of the nucleons.\
The spin of an excited MNT product is also affected by the evaporation of nucleons, although this effect was not included.\

In the second step, the IR of $^{211}$Po has been calculated by splitting the spin distribution into two parts:\ the lower spin distribution region is assumed to feed the ground-state (9/2$^+$), and the higher-spin distribution to the isomeric state (25/2$^+$).\ 
The spread in the spin distribution can be anticipated due to the evaporation of nucleons and the cascade of $\gamma$-decays from the MNT product.\ To account for these effects, an empirical systematic approach was applied for the calculation of IR using equation (1), consisting of an effective angular momentum cutoff $J_{\rm eff}$, and a spreading parameter $\Delta$ \cite{Gasques06}, 
\begin{equation}
	\label{mmeq1}
	IR=\frac{\sum_{J} Y_{J}^{(\rm m)}}{\sum_{J} Y_{J}^{(\rm g)}},
\end{equation} 
\begin{equation}
	\label{mmeq2}
	Y_{J}^{(\rm g)}=\frac{Y_{J}^{(\rm theory)}}{(1+exp{\frac{(J_{\rm eff} - J)}{\Delta}})};
	Y_{J}^{(\rm m)}=\frac{Y_{J}^{(\rm theory)}}{(1+exp{\frac{(J - J_{\rm eff})}{\Delta}})},
\end{equation}

The $Y_{J}^{(\rm theory)}$ corresponds to the theoretical estimation of spin distribution for a particular MNT product, and $Y_{J}^{(\rm m)}$ ($Y_{J}^{(\rm g)}$) represents the spin distribution associated with the ground-state (isomeric state).\ 
In our calculations, we assumed $\Delta$ = 0.5, which is justified to account for the angular momentum carried away by the neutrons and $\gamma$ rays \cite{Gasques06,Hafner19,Pal08}.

In this work, the spin distribution of $^{211}$Po was calculated for both reactions:\ $^{136}$Xe+$^{209}$Bi at E/V${\rm _B}$=1.12, 1.26, and 1.33; and $^{136}$Xe+$^{\rm nat}$Pb at E/V${\rm _B}$=1.13, 1.27, and 1.34.\
Fig.\ \ref{fig3}(a) demonstrates the variation in the spin distributions of $^{211}$Po for the two energies in both cases.\ It is found that the most probable value of spin distributions at near barrier energy ($i.e.$, E/V${\rm _B}$=1.12 and 1.13) is lower compared to the spin of the isomeric state of $^{211}$Po (25/2$^+$) as marked using a vertical line.\ 
However, the spin distributions for higher projectile energy (at E/V${\rm _B}$=1.33 and 1.34) are significantly different for both reactions.\ The most probable value of the spin distribution obtained from $^{136}$Xe+$^{\rm nat}$Pb and $^{136}$Xe+$^{209}$Bi is significantly larger and lower compared to the spin of $^{211}$Po isomer, respectively.\ 
Variation in the spin distributions considering with and without the angular acceptance of an MNT gas cell can be seen in Fig.\ \ref{fig3}(b).\ Feeding of the spin distribution into the ground and isomeric states estimated from equations (1) and (2) is represented with a solid line for the angular acceptance 22$^{\circ}$-55$^{\circ}$ for E/V${\rm _B}$=1.33.\

\section{\label{s4} Discussions}
The $^{209}$Bi target differs from the $^{\rm nat}$Pb target in terms of relatively high ground-state spin of $^{209}$Bi (9/2$^-$) compared to almost zero spin of $^{\rm nat}$Pb (only $^{207}$Pb, with 22.1\% isotopic abundance, has a non-zero spin 1/2$^-$).\ The ground-state spin of $^{211}$Po has the identical spin as $^{209}$Bi but with opposite parity (9/2$^+$).\ 
Theory suggests that a larger production of high spin isomers is more probable in heavy-ion-induced MNT reactions due to the large angular momentum brought by projectiles compared to other conventional nuclear reaction processes.\ This is clearly endorsed by the observed $\alpha$-spectra in which population of $^{\rm 211m}$Po, $^{\rm 212m}$At dominates over corresponding ground states.\

\begin{figure}[t]
	\centering
	\includegraphics[width=84mm,height=108mm]{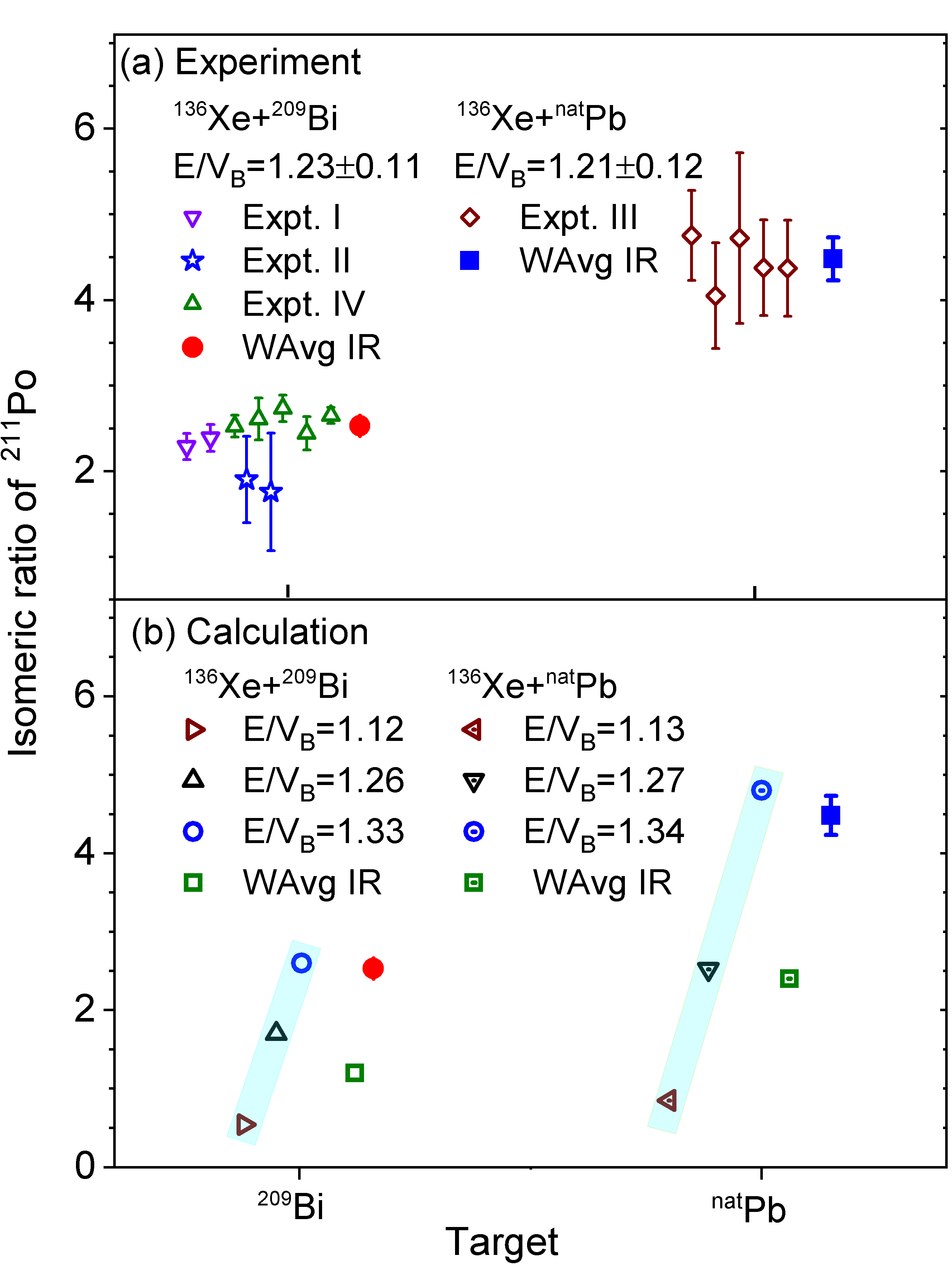}
	\caption{\label{fig4}(Color online)\ (a) IRs of $^{211}$Po produced in $^{136}$Xe+$^{209}$Bi at E/V${\rm _B}$=1.23$\pm$0.11 and $^{136}$Xe+$^{\rm nat}$Pb at E/V${\rm _B}$=1.21$\pm$0.12 corresponding to 945 MeV beam energy in different experiments.\ Details of the experimental parameters are described in Table\ 1.\ (b) Estimation of IRs of $^{211}$Po at three projectile energies and comparison of WAvg of IRs with measured results for $^{136}$Xe+$^{209}$Bi and $^{136}$Xe+$^{\rm nat}$Pb.\ WAvg refers to the Weighted Average.\ Strip lines are shown as a guide to the eye for increasing IRs with projectile energy.}
\end{figure}

As expected, one of the crucial observations reflected in Fig.\ \ref{fig4}(a) (shown with open symbols) is the IRs of $^{211}$Po against various experimental parameters measured in the different experiments of $^{136}$Xe+$^{209}$Bi and $^{136}$Xe+$^{\rm nat}$Pb.\ Different observations were found to be in excellent agreement.\
The final value of IR was determined by considering the weighted average (WAvg) of the measured IRs, shown with solid symbols.\
It should be stressed here that the IR of $^{211}$Po for $^{136}$Xe+$^{\rm nat}$Pb is found to be $\approx$1.8 times higher in comparison to $^{136}$Xe+$^{209}$Bi.\
Possible reasons for the different IR at an incident energy of  945 MeV are:\ (i) the intrinsic spin of the target, (ii) the angular momentum brought in by the projectile's momentum, and (iii) the number of transferred nucleons.\ It must be noted that the angular momentum brought in due to the momentum of 945 MeV $^{136}$Xe beam in the formation of $^{211}$Po from both reactions will be identical; however, coupling with the intrinsic spin of different targets could result in different spin values.\

Different feasible production channels of $^{211}$Po from $^{136}$Xe+$^{\rm nat}$Pb reaction would be 2$p$1$n$, 2$p$2$n$, 2$p$3$n$ corresponding to dominant isotopic abundance of $^{208}$Po (52.4\%), $^{207}$Po (22.1\%), and $^{206}$Po (24.1\%).\ However, it is important to notice that production of $^{211}$Po from the 2$p$2$n$, 2$p$3$n$ channels would not only be suppressed by isotopic abundances but also from the significant reduction of MNT cross-sections with increasing number of transfer channels \cite{Karpov17}.\ Therefore, 2$p$2$n$ channel would only be an effectively dominant route in the population of $^{211}$Po.\ However, in case of $^{136}$Xe+$^{\rm 209}$Bi, the production route $^{211}$Po would be 2$p$1$n$ channel.\ Moreover, it should be noted that the effect of 1$n$-transfer would be the same in both cases.\ Hence, 1$p$ and 2$p$ transfer can effectively be responsible for any significant variation in the production of $^{211}$Po for both reactions.\

A comparison of measured and computed IRs is represented in Fig.\ \ref{fig4}(b).\ The theoretical calculation of IRs at 12-13\% above the barrier (i.e., IR$<$1) clearly indicates the smaller production of isomer compared to the ground state of $^{211}$Po.\ Additionally, the IRs have similar values for both reactions despite the different spin of the targets and production routes.\ This means at near barrier energy, the coupling of the intrinsic spin of $^{209}$Bi and the spin brought in by the projectile as well as the spin transferred to the $^{209}$Bi by 1$p$-transfer is equivalent to the coupling of spin due to the projectile and spin transferred to the $^{\rm nat}$Pb by the $2p$-transfer.\ However, the influence of the target spin on the production of MNT fragments will be more probable to appear at near-barrier energies due to the minimum amount of angular momentum that the projectile brings.\ Thus, the calculation implies a weak dependence of the spin distribution of $^{211}$Po on the spin of $^{209}$Bi and $^{\rm nat}$Pb targets at near-barrier energy.\ Moreover, target spin dependence on the spin distribution of an MNT fragment would hardly appear at higher projectile energies.\

In figure \ref{fig4}(a) (solid symbols), the quantitative deduction of IRs at 945 MeV bombarding energy (i.e., IR$>$1) indicates that the isomer production of $^{211}$Po significantly dominates over the ground state.\ The calculated IRs shown in Fig. \ref{fig4}(b) were found to increase with increasing projectile energy for both reactions.\ 
The increasing nature of IR can be understood as the projectile energy could bring in more angular momentum to the system via the projectile's momentum as well as the transfer of nucleons (in the case of MNT-induced reactions), indicating a large probability of the population of high spin states ($i.e.$, isomers, in the present case) of MNT products and thereby resulting in higher values of IRs.\
The optimum angular range of the new gas cells was considered in the theoretical calculation to match the experimental and theoretical scenario.\ Finally, the weighted average (WAvg) of the estimated IRs was computed to simulate the production of the isomeric (25/2$^+$) and ground (9/2$^+$) states of $^{211}$Po for the projectile energy loss within the targets.\

It is evident from Fig.\ \ref{fig4}(b) that the population of $^{211}$Po isomer over the ground state differs for the studied production routes, $^{136}$Xe+$^{\rm nat}$Pb and $^{136}$Xe+$^{209}$Bi.\ Theoretical calculations (open squares) qualitatively explain the measured values.\ However, it underpredicts the experimental IRs by a factor of two for both reactions.\ Tentative assignment of (25/2$^+$) of $^{\rm 211m}$Po might be one reason for the quantitative disagreement.\ The tentative assignment was made using empirical shell-model (ESM) calculations \cite{TR98}.\ Moreover, decay of the other two isomeric states (31/2$^-$) and (43/2$^+$) of $^{211}$Po (tentatively assigned spin values) might change the independent production of ground (9/2$^+$) and/or isomeric (25/2$^+$) states, which could result in this inconsistency.\ Additionally, as discussed in Sec.\ \ref{s3}, the angular momentum of TLFs would not be just a sum of the ground-state spin of the target and the angular momentum transferred by projectiles and nucleon transfer.\ One must also consider the angular momentum carried away by nucleon evaporation from the excited MNT products, which would influence the final spin of the product.\ However, it was not considered in the calculations.\ More experimental data for different target-projectile systems would be helpful to properly incorporate its contribution in the theoretical calculation.\ Nonetheless, it is worth concluding that the $p$-transfer channels are strongly correlated with spin distributions and, thereby, the IRs.\ This means that more $p$-transfer could impart more spin to the MNT fragments and, therefore, more significant production of high spin state isomers.\

\begin{figure} [t]
	\centering
	\includegraphics[width=84mm,height=62mm]{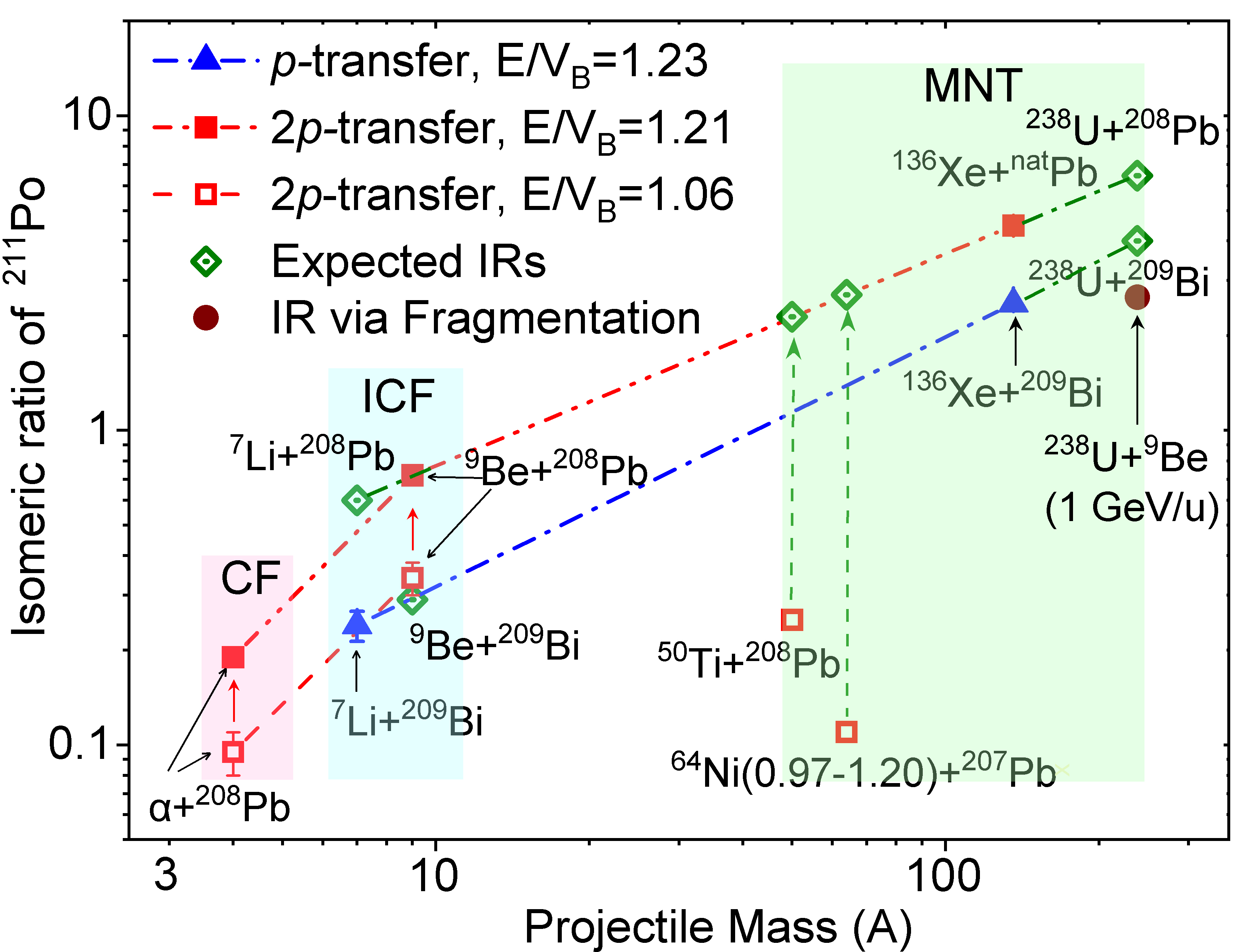}
	\caption{\label{fig5}(Color online)\ Comparison of different experimental IRs of $^{211}$Po with different projectile masses corresponding to various nuclear reaction processes:\ $\alpha$+$^{208}$Pb\cite{Barnett74}, $^7$Li+$^{209}$Bi and $^{9}$Be+$^{208}$Pb\cite{Gasques06}, $^{50}$Ti+$^{208}$Pb \cite{Devraja22}, $^{64}$Ni +$^{207}$Pb \cite{Comas13}, $^{136}$Xe+$^{209}$Bi/$^{\rm nat}$Pb (present work) at 21-23\% above the barrier, and $^{238}$U+$^{9}$Be at 1 GeV/u in fragmentation process \cite{Timo15}.}
\end{figure} 

Figure\ \ref{fig5} exhibits a comparative study on the IRs of $^{211}$Po deduced in the present work together with other experimental results over a wide range of projectile masses.\ The different shaded regions represent distinct nuclear reaction processes:\ (i) complete fusion (CF) process in $\alpha$+$^{208}$Pb (solid square)\cite{Barnett74}; (ii) incomplete fusion (ICF) process involved in $^{7}$Li +$^{209}$Bi (solid triangle) and $^{9}$Be+$^{208}$Pb (solid square) \cite{Gasques06}; and (iii) MNT reaction processes using $^{50}$Ti+$^{208}$Pb (open square), $^{64}$Ni+$^{207}$Pb (open square), $^{136}$Xe+$^{nat}$Pb (solid square), $^{136}$Xe+$^{209}$Bi (solid triangle) within 21-23\% above barrier energies \cite{Comas13,Devraja22}, and (iv) via fragmentation reaction process using $^{238}$U+$^{9}$Be at 1 GeV/u (solid circle) \cite{Timo15}.\ The IR of $\alpha$+$^{208}$Pb and $^{9}$Be+$^{208}$Pb at E/V${\rm _B}$=1.21 were extrapolated from the increasing values of IRs reported at lower incident energies; one of them is shown with the open square at E/V${\rm _B}$=1.06.\ However, an increment in the IRs and subsequent decrement may be anticipated with increasing projectile energies, similar to other reactions \cite{KumarEPJP21,Rinku21}.\ Therefore, the reactions are essential to validate experimentally.\ Nonetheless, analysis endorses the increment of IRs populated from the $2p$-channel compared to the 1$p$-channel.\ Anticipated IRs of $^{211}$Po for several reactions were shown with an open diamond symbol.\ The 1$p$- and $2p$-transfer channels were indicated with dash-dotted and dash-dot-dot lines, respectively.\ 

The comparative analysis demonstrates that the IRs of $^{211}$Po have primarily been affected by two entrance channel parameters:\ the projectile mass and the transfer channel production route.\ It is apparent that IRs of $^{211}$Po gradually increase with the projectile mass, which reflects the sensitivity of spin distribution of the reaction products brought in by the mass of projectiles in different reaction processes (see Fig.\ \ref{fig5}).\ The ICF process is similar to transfer-like processes in which cluster transfer can be favored due to the weakly bound nature of projectiles, like $^{6,7}$Li, $^{9}$Be \cite{Kumar17,Prajapat21,Deepak17}.\
In the ICF process, an enhanced angular momentum transfer has been observed in evaporation residues (ERs) compared to CF corresponding to the same production channels; e.g., $\alpha$-transfer from $^7$Li in ICF would impart more spin to ER compared to $\alpha$ particle CF \cite{Gasques06}.\ Similarly, in the present case, the residue produced from the MNT-induced reaction process via either 1$p$- or 2$p$-transfer channel can impart more angular momentum to the system than the one formed from the identical transfer of nucleons in the ICF process \cite{Gasques06}.\ This indicates that spin transferred via nucleon transfer strongly correlates with the projectile mass.\ Additionally, it is worth noticing the consistent enhancement of IRs for the $2p$-transfer channel compared to the 1$p$-transfer channel over an extended mass region in distinct CF, ICF, and MNT reaction processes.\  

The IRs of $^{211}$Po produced from $^{50}$Ti+$^{208}$Pb and $^{64}$Ni+$^{207}$Pb reactions were deduced from the $\alpha$-spectra at relatively lower projectile energy, E/V${\rm _B}$=1.06 and E/V${\rm _B}$=0.21-1.20, respectively.\ The experiments were not aimed for the investigation of IRs \cite{Devraja22,Comas13}.\ Significantly lower values of the IR were found for both reactions shown in Fig.\ \ref{fig5}.\ This might primarily be due to lower projectile energy or due to the limited angular coverage of the experimental setup (0$^\circ$$\pm$2$^\circ$) \cite{Liao23}.\ 
The trend line obtained from the IRs of $^{211}$Po from ICF and MNT processes predicts the large value of IRs for $^{50}$Ti and $^{64}$Ni projectiles at E/V${\rm _B}$=1.21.\ Moreover, the IRs from $^{238}$U+$^{209}$Bi/$^{\rm nat}$Pb are predicted to be large compared to one obtained from $^{238}$U+$^{9}$Be via fragmentation process.\ Hence, comparative analysis prompts the investigation of IRs for different target-projectile combinations populating $^{211}$Po, including MNT reactions: $^{50}$Ti+$^{208}$Pb, $^{64}$Ni+$^{207}$Pb, and $^{238}$U+$^{208}$Pb/$^{209}$Bi.\ Examining the aforementioned reactions at near-barrier energies would be worthwhile for the limpid perspicacity of the spin distributions.\ That is why exploring the $^{238}$U+$^{209}$Bi/$^{208}$Pb/$^{238}$U reactions using the MNT approach is one of the prime objectives of an approved proposal to be performed soon at FRS-IC, GSI \cite{TimoGSI}.\

\section{\label{s6}Summary}
First measurement on the IRs of $^{211}$Po was accomplished from the $\alpha$-decay spectra produced via different channels of MNT reactions using $^{136}$Xe+$^{209}$Bi and $^{136}$Xe+$^{\rm nat}$Pb.\ The population of isomers over the corresponding ground states was dominantly observed in different measurements of $\alpha$-spectra.\ A dynamical approach based on Langevin equations was utilized to compute spin distributions of the MNT fragment for the first time and subsequently estimate the IRs at three distinct energies for both reactions.\
Close agreement between the computed IRs of $^{211}$Po from $^{136}$Xe+$^{209}$Bi and $^{136}$Xe+$^{\rm nat}$Pb reactions at near-barrier energy indicate a weak dependence of target spin on the spin distribution and thereby would hardly affect the IRs at high energies, $i.e.$, 26-34\% above the barrier.\ Deduced IRs of $^{211}$Po from $^{136}$Xe+$^{\rm nat}$Pb has been found to be increased by a factor of $\approx$1.8-times than obtained from $^{136}$Xe+$^{209}$Bi.\ The considerable increment in $^{211}$Po isomer can be attributed to the production route of the 2$p$-transfer channel in $^{136}$Xe+$^{\rm nat}$Pb compared to the 1$p$-transfer channel in $^{136}$Xe+$^{209}$Bi.\ The estimated IRs were found to be strongly affected by projectile energy and qualitatively consistent with experimental findings of both reactions.\ However, theoretical estimations underpredict the measured IRs by a factor of two.\

Comparative analysis on the IRs of $^{211}$Po over the projectile mass in different nuclear reaction processes at E/V${\rm _B}$$\approx$1.21-1.23 reveals two main entrance channel parameters:\ projectile mass and transfer channel production route, which strongly affect the IRs and, thereby, would play a major role in the spin distributions.\ The IRs of $^{211}$Po has been found to be enhanced for the 2$p$-channel than for the 1$p$-channel over an extended mass range of the projectiles, inducing via CF, ICF, and MNT reaction processes.\ Present experimental and theoretical findings invoke for the comprehensive and systematic works to validate the predicted IRs for $^{9}$Be+$^{209}$Bi, $^{7}$Li+$^{208}$Pb, $^{50}$Ti/$^{64}$Ni+$^{208}$Pb, $^{238}$U+$^{208}$Pb/$^{209}$Bi including many other feasible target-projectile systems above Coulomb barrier energies.\

\section*{Acknowledgements}
The author would like to express his sincere and deep sense of indebtedness to Dr.\ A.V. Karpov and Dr.\ V.V.\ Saiko for their unconditional continuous support in the theoretical estimation works.\
We thank the cyclotron team for their efforts to provide a stable beam condition.\ The work was financially supported by the EU Horizon 2020 research and innovation program (ERC Consolidator Grant 2017) under Grant Agreement No.\ 771036, 861198–LISA–H2020-MSCA-ITN-2019, and from DAAD Grant No.\ 57610603.\ Zs.P. acknowledges support by the Science and Technologies Facilities Council (STFC).\ P.C. acknowledges contract PN 23.21.01.06 of the Romanian Ministry of Research and Innovation.\ I.M. acknowledges partial support by the Israel Science Foundation, Grant No. 2575/21.\ 


\bibliographystyle{elsarticle-harv} 

\end{document}